\documentclass[conference]{IEEEtran}
\IEEEoverridecommandlockouts
\usepackage{cite}
\usepackage{amsmath,amssymb,amsfonts}
\usepackage{algorithmic}
\usepackage{graphicx}
\usepackage{textcomp}
\usepackage{xcolor}
\def\BibTeX{{\rm B\kern-.05em{\sc i\kern-.025em b}\kern-.08em
    T\kern-.1667em\lower.7ex\hbox{E}\kern-.125emX}}

\usepackage[utf8]{inputenc}
\usepackage[T1]{fontenc}
\usepackage{url}
\usepackage{ifthen}
\usepackage{cite}
\usepackage{stfloats}
\usepackage{epstopdf}
\usepackage{amsfonts}
\usepackage{enumitem}
\usepackage{amssymb}
\usepackage{mathrsfs}
\usepackage{bm}
\usepackage{mdwmath}
\usepackage{subfigure}
\usepackage{mdwtab}
\usepackage{dsfont}
\usepackage{color}
\usepackage{verbatim}
\usepackage{array}
\usepackage{epstopdf}
\usepackage{multirow}
\usepackage{booktabs}
\usepackage{lineno}
\usepackage{amsmath}
\usepackage{makecell}
\usepackage{algorithmic}
\usepackage{amsthm}
\usepackage{fancyhdr}
\usepackage{flushend}
\usepackage[ruled,vlined,linesnumbered]{algorithm2e}

\allowdisplaybreaks

\usepackage{titletoc}

\renewcommand\thesubsubsection{\alph{subsubsection}}

\begin{document}
\title{Real-time Extended Reality Video Transmission Optimization Based on Frame-priority Scheduling}


\author{Guangjin Pan$^{\dagger}$, Shugong Xu$^{\dagger}$, Shunqing Zhang$^{\dagger}$, Xiaojing Chen$^{\dagger}$, and Yanzan Sun$^{\dagger}$\\
$^{\dagger}$
School of Communication and Information Engineering, Shanghai University, Shanghai 200444, China.\\
Email: \{guangjin\_pan, shugong, shunqing, jodiechen, yanzansun\}@shu.edu.cn\\
\thanks{ Shugong Xu is the corresponding author.}
}

%



\IEEEtitleabstractindextext{%
\begin{abstract}
Extended reality (XR) is one of the most important applications of 5G. For real-time XR video transmission in 5G networks, a low latency and high data rate are required. In this paper, we propose a resource allocation scheme based on frame-priority scheduling to meet these requirements. The optimization problem is modelled as a frame-priority-based radio resource scheduling problem to improve transmission quality. We propose a scheduling framework based on multi-step Deep Q-network (MS-DQN) and design a neural network model based on convolutional neural network (CNN). Simulation results show that the scheduling framework based on frame-priority and MS-DQN can improve transmission quality by 49.9\%-80.2\%.
\end{abstract}

}

\maketitle

\IEEEdisplaynontitleabstractindextext

%
\IEEEpeerreviewmaketitle


\section{Introduction}
\IEEEPARstart{R}{ecently}, Metaverse \cite{Metaverse} integrates virtual and real environments, providing an immersive experience that sparks unlimited imagination for the future. In order to achieve the metaverse's vision, eXtended Reality (XR) technology is undergoing rapid development. XR encompasses various types of reality, including Augmented Reality (AR), Virtual Reality (VR), Mixed Reality (MR) and Cloud Gaming (CG) \cite{cite:TS26928}, all of which require low-latency transmission and interaction. Typically, XR services are rendered on the cloud servers and then transmitted to the XR clients for playback. To ensure real-time XR content, real-time XR videos are usually transmitted frame by frame. XR services fall under the category of Real-Time Broadband Communication (RTBC) services. To ensure a seamless, high-quality XR video experience, the system must fulfill the standards of low latency and high data rate. 


Most wireless resource allocation algorithms often assign wireless resources to users based on their weights. However, even for the same user, the importance weight may vary for different information. In particular, within the XR transmission system, fulfilling the dual requirements of high bandwidth and low latency inevitably leads to the failure of certain information transmission. This prompts us to explore obtaining the importance of different information in XR videos in order to determine transmission priorities and ensure the transmission of important information. For example, in the Group of Picture (GOP) based XR video encoding model \cite{cite:TS38838}, I-frames are usually more important than P-frames. The XR server can also determine the importance of video frames according to other criteria and inform the BS in some way to assist in scheduling. In conclusion, the higher the importance of successfully transmitting XR video information, the higher the transmission quality as well.


Several related works have been developed on wireless resource scheduling, as discussed in \cite{PF,cite:exprule,cite:M-EDF-PF,cite:priority-DDPG,cite:RPPO,cite:frameintegrated}. The proportional fair scheduler was analyzed in \cite{PF} to obtain the cell throughput, while scheduling algorithms like Exponential Rule (EXP-RULE) \cite{cite:exprule} and Modified Earliest Deadline First and Proportion Fair (M-EDF-PF) \cite{cite:M-EDF-PF} were proposed to support real-time traffics. In \cite{cite:priority-DDPG}, the Deep Deterministic Policy Gradient (DDPG) algorithm was utilized to minimize queuing delay experienced by devices. Additionally, a delay-oriented scheduling algorithm based on partially observable Markov decision process was proposed in \cite{cite:RPPO} to reduce the tail delay and average delay. However, these works do not consider the traffic characteristics of XR. \cite{cite:frameintegrated} focused on frame-level integrated transmission to enhance the frame success rate. However, \cite{cite:frameintegrated} doesn't consider the importance of video frames.

\begin{figure}[tb]
\centering 
\includegraphics[height=2.1in,width=3.5in]{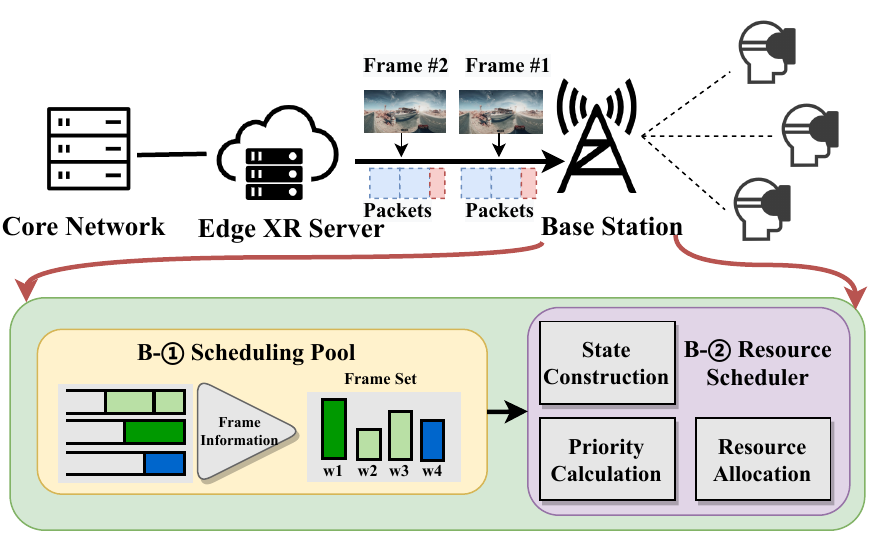}
\caption{XR transmission system framework. Real-time XR video is rendered on edge XR servers and then transmitted to clients for playback. BS uses frame information to schedule packets and improve transmission quality.}
\label{fig:framework} 
\end{figure}

As shown in Fig. \ref{fig:framework}, we consider a real-time XR oriented wireless transmission system. The edge XR server performs real-time video encoding and transmits each encoded video frame to the base station (BS) with each frame consisting of a set of video packets. Subsequently, the BS schedules the video packets in the queue and delivers them to XR devices, which then decode and play the video. The contributions of our study are listed below,
\begin{itemize}
  \item{\textit{Frame-priority-based wireless resource scheduling problem formulation.}} In this paper, our main aim is to improve the transmission quality of the real-time XR video. We model the transmission quality as the sum of the importance weights of the frames that failed to transmit. In summary, we believe that a decrease in the total sum of importance weights for frames that were not successfully transmitted leads to improved transmission quality. To enhance the transmission quality, we adopt a scheduling scheme based on frame priority and perform traffic scheduling frame-by-frame. Based on the frame priority scheduling strategy, we model the problem as an optimization problem of frame priority and maximize the transmission quality.
  \item{\textit{MS-DQN-based wireless resource scheduling scheme.}} To solve the proposed problem, we propose a multi-step deep Q-network (MS-DQN) algorithm to reduce the action space, allowing the agent to make multi-step decisions within each time slot to allocate radio resources. Meanwhile, we have designed a convolutional neural network (CNN) based action network to deal with the dynamically changing action space and state space. Compared to the baseline algorithm, our proposed algorithm improves the transmission quality by over 49.9\%.
\end{itemize}

The rest of this paper is organized as follows. In Section \ref{Sec2}, we introduce system models and formulate the frame-priority-based optimization problem. Section \ref{Sec4} gives the proposed framework and MS-DQN based scheduling scheme. Numerical results and analysis are presented in Section \ref{Sec5}. Finally, the paper is concluded in Section \ref{Sec6}.

\section{System Model} \label{Sec2}

In this section, we consider an XR video 5G transmission system with one edge XR server, one BS, and $N$ XR devices denoted by the set $\mathcal{N} = \{1, 2, \cdots, N \}$. The time is assumed to be discretized into several slots $ \mathcal{T}=\{1,2,\cdots,T\}$. The length of one slot $\Delta t$ is equal to the time length of one transmission time interval (TTI) of the 5G system. Therefore, the BS allocates resource blocks (RBs) once in each time slot $t$.

\subsection{XR Traffic Model}

XR videos typically generate video frames at regular intervals based on a given frame rate. Let $\mathcal{K}_n$ denote the frame set of device $n$  transmitted in a unit of time. For each frame ($k \in  \mathcal{K}_{n}$), the frame size is $r_{n,k}$. For each user $n$, the arrival time of the $k$-th frame at the BS can be modelled as,
\begin{eqnarray}\label{equ2-A-1}  
t_{n,k}= t_{n,0}+\lfloor \frac{\frac{k-1}{f_n}+t_{jitter}}{\Delta t} \rfloor,
\end{eqnarray}
where $t_{n,0}$ is the initial arrival time, $f_n$ is the frame rate, and $t_{jitter}$ is the arrival jitter time due to the codec and transmission. The value of $t_{jitter}$ is a random variable that follows a certain distribution.

In addition, considering the different types of frames and encoding methods, different XR video frames have varying levels of importance. For example, I-frames contain all the information of a video frame and can be decoded independently, while P-frames require decoding the information from the previous frame. Therefore, compared to P-frames, I-frames have more data and are more important. We define the importance weight of frame $k$ for device $n$ as $w_{n,k}$.

\subsection{Channel and Scheduling Models} 
Considering there are $N^{rb}_t$ available RBs in time slot $t$, for each RB, the achievable data rate of device $n$ at the $t$-th slot is given by,
\begin{eqnarray}\label{equ2-B-1}  
{c}_{n,t}= \Delta t B_w \log(1+\frac{p_n h_{n,t}}{B_w\sigma^2}),
\end{eqnarray}
where $p_n$ is the transmit power of device $n$, $h_{n,t}$ is the channel gain of device $n$ in time slot $t$,  $B_w$ is the bandwidth of each RB, $\sigma^2$ is the Gaussian noise power spectral density. 

In the XR video transmission task, it is essential to fully transmit packets belonging to the same frame before the frame delay budget (FDB) $t^{\star}$ \cite{cite:frameintegrated}. Let $\widetilde{r}_{n,k,t}$ denote the remaining data size of frame $k$ waiting for transmission in time slot $t$. It can be given by,
\begin{eqnarray}\label{equ2-B-2}
\widetilde{r}_{n,k,t} = r_{n,k} - \sum_{\tau=t_{n,k}}^{t-1} N_{n,k,\tau}^{rb} {c}_{n,\tau},
\end{eqnarray}
where $r_{n,k}$ represents the total size of the frame, $N_{n,k,t}^{rb}$ denotes the number of RBs allocated to device $n$ for frame $k$ in time slot $t$, and ${c}_{n,\tau}$ represents the achievable data rate of device $n$ in RBs at time slot $\tau$, as calculated in Equation~\eqref{equ2-B-1}.

Therefore, a frame can be considered as successfully transmitted if the condition
\begin{eqnarray}\label{equ2-B-3}
x_{n,k} = \mathbb{I}(\widetilde{r}_{n,k,t_{n,k}+t^{\star}} \le 0) = 1
\end{eqnarray}
is satisfied. Otherwise, $x_{n,k} =0$.

To enhance the transmission capability of XR in wireless networks, we propose that the BS extracts some frame information from the real-time transport protocol (RTP) header\footnote{Although the current RTP standard does not support this assumption, we believe that if the BS can obtain some frame information, the transmission quality of XR video can be effectively improved.}. By utilizing this information, the BS can sort the packets waiting to be transmitted and schedule them accordingly. Specifically, the set of frames waiting to be transmitted during time slot $t$ is denoted by $\mathcal{K}_{t}= \{(n,k)|n \in \mathcal{N}, k \in \mathcal{K}_{n,t} \}$, where $\mathcal{K}_{n,t}=\{k|k\in \mathcal{K}_n,0 \le t-t_{n,k} < t^{\star}, \widetilde{r}_{n,k,t}>0\}$ represents the set of frames for device $n$ that are waiting to be transmitted in time slot $t$.

The BS can perform scheduling based on the priority of frames, which is determined by the set of frames received. Let $p_{n,k,t}$ denote the priority of device $n$ for frame $k$ in time slot $t$. The resource allocation policy can be expressed as,
\begin{eqnarray}\label{equ3-A-2}  
N_{n,k,t}^{rb}=\!  \left\{
{\begin{aligned}
\lceil \frac{\widetilde{r}_{n,k,t}}{{c}_{n,t}} \rceil   ,   \mathop{\textrm{if}} \widetilde{r}_{n,k,t}<{c}_{n,t}(N_{t}^{rb}- \! \! \! \! \! \! \! \! \! \! \! \sum_{(n',k')\in \mathcal{K}_{n,k,t}} \! \! \! \!  \! \! \! \! \! \! \!  N_{n',k',t}^{rb}),\\
N_{t}^{rb}- \! \! \! \! \! \! \! \! \! \! \! \sum_{(n',k')\in \mathcal{K}_{n,k,t}} \! \! \! \! \! \! \! \! \! N_{n',k',t}^{rb} \qquad \qquad \quad \ \ , \mathop{\textrm{otherwise}} ,
\end{aligned}}  \right. \label{equ:allocation}
\end{eqnarray}
where $\mathcal{K}_{n,k,t}=\{(n',k') | (n',k') \in \mathcal{K}_t, \ p_{n,k,t}<p_{n',k',t}\}$ is the set of frames with a priority higher than the frame $(n,k)$.

As mentioned earlier, the traditional flow-priority-based scheduling policy primarily considers the device's queue lengths (to ensure fairness) and the channel state information (to optimize throughput) \cite{cite:RPPO}. However, in the frame-priority-based scheduler, the priority is calculated by,
\begin{eqnarray}
\bm{p}_{t} &=&{\pi}^p(\bm{{r}}_{t},\bm{\widetilde{r}}_{t},\bm{\tau}_t,\bm{h}_t),
\label{equ:priority}
\end{eqnarray}
where $\bm{p}_{t}\triangleq \{{p}_{n,k,t},\forall (n,k) \in \mathcal{K}_{t}\}$, $\bm{{r}}_t\triangleq\{{r}_{n,k},\forall (n,k) \in \mathcal{K}_{t}\}$, $\bm{\widetilde{r}}_t\triangleq\{\widetilde{r}_{n,k,t},\forall (n,k) \in \mathcal{K}_{t}\}$, $\bm{\tau}_t\triangleq\{\tau_{n,k,t},\forall (n,k) \in \mathcal{K}_{t}\}$, and $\bm{h}_t\triangleq\{h_{n,t},\forall (n) \in \mathcal{N}\}$. Specifically, $\tau_{n,k,t}=t^{\star}+t_{n,k}-t$ is the remaining frame delay budget (RFDB). ${\pi}^p(\cdot)$ represents the scheduler policy.

\subsection{Problem Formulation} \label{Sec3}
In order to describe the relationship between video experience quality and scheduling scheme, we define video transmission quality as the sum of the weights of frames that failed to transmit within a unit of time.
\begin{eqnarray}
Q_n &=&- \sum_{k \in \mathcal{K}_n } (1-x_{n,k})w_{n,k}.\label{equ:qoe_t}
\end{eqnarray}

In this paper, we aim to find a wireless resource allocation scheme ${{\pi}^p}(\cdot)$ to maximize the transmission quality. The original problem can be formulated as,

Problem $\mathrm{P} 1$ (\textit{Original Problem}):
\begin{eqnarray} \mathop{\textrm{maximize}}_{{{\pi}^p}(\cdot)} && \sum_{n \in \mathcal{N} } \mathbb{E}[Q_n] \label{equ-OriginalProblem} \\
\mathop{\textrm{s.t.}} 
&& \! \! \! \! \sum_{(n,k) \in \mathcal{K}_t}N_{n,k,t}^{rb} \le N_t, \forall{t}, \\
&& \! \! \! \! \eqref{equ:allocation},\eqref{equ:priority},\eqref{equ:qoe_t}. \nonumber
\end{eqnarray}

The above problem is generally difficult to solve due to the following reasons. Firstly, in the context of long-term optimization, the dynamic changes in the network environment, including the wireless channel and available network resources, increase the complexity of the problem. Additionally, the uncertainty in the frame size and arrival time poses challenges to the effective solution of the problem. It is difficult to obtain an optimal closed-form expression for the priority-based scheduling scheme through mathematical analysis. Therefore, we propose an MS-DQN based solution to optimize the transmission quality of real-time XR video.

\section{frame-priority-based scheduling method} \label{Sec4}

\begin{figure}[tb]
\centering 
\includegraphics[height=2.1in,width=3.5in]{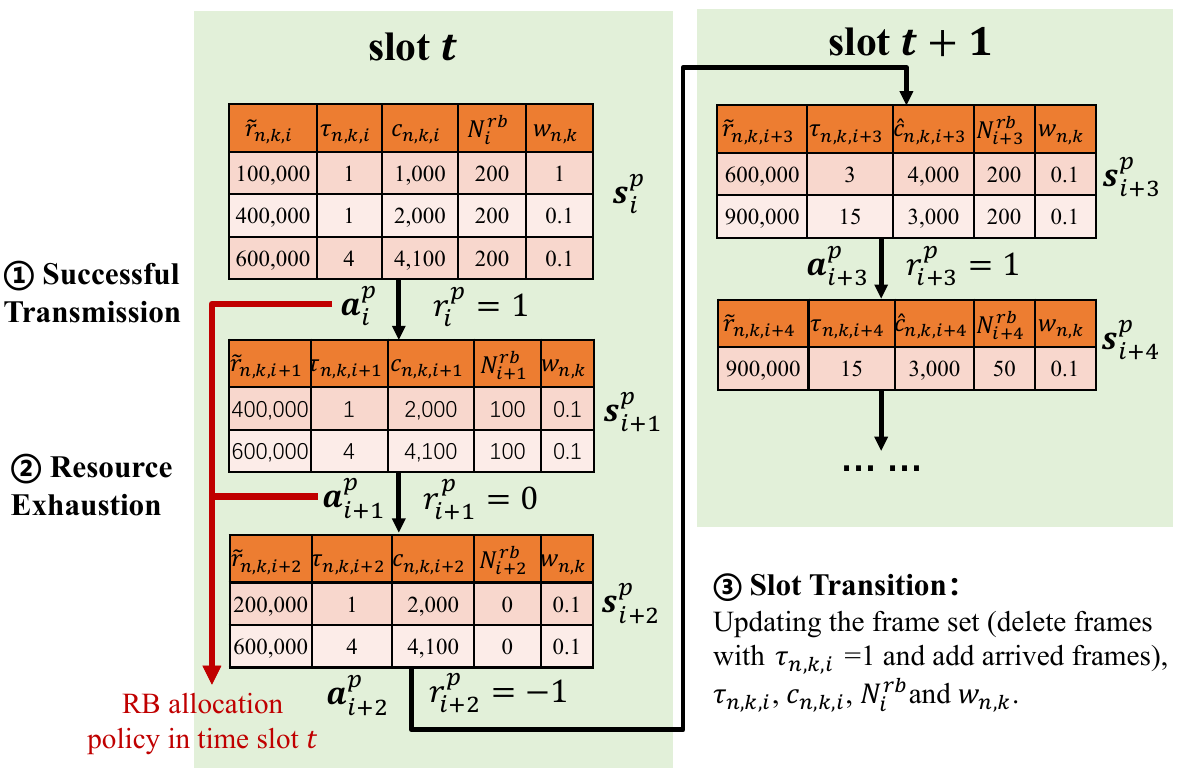}
\caption{This example demonstrates three types of state transitions. In slot $t$, the agent of MS-DQN makes two scheduling decisions and transmits two frames.}
\label{fig-priority} 
\vspace{-3mm}
\end{figure}

In a real-time XR transmission system, adopting an appropriate radio resource scheduling scheme can effectively enhance the transmission quality. In this section, we propose the MS-DQN based resource scheduling framework to solve the problem. When the action space is discrete and finite, DQN can work well. However, in the problem $\mathrm{P} 1$, the frame priority $p_{n,k,t}$ is a continuous value, and the action space is infinite. Although we can sort the priorities to select the scheduled users and discretize the action, the action space is still large. To reduce the action space, we proposed a multi-step based DQN algorithm. In our proposed MS-DQN algorithm, at each decision step, the agent outputs the probability values associated with each frame scheduling, which can be regarded as the priority of each frame. The proposed algorithm allows the agent to make multi-step decisions in each slot and select one frame of data to transmit in each decision step. Fig. \ref{fig-priority} shows an example of the state transition. In this example, the agent makes a three-step decision in slot $t$. Then the BS allocates RBs based on actions $\bm{a}^p_i$, $\bm{a}^p_{i+1}$, and $\bm{a}^p_{i+2}$. Then, the state transitions to $\bm{a}^p_{i+3}$, which is in slot $t+1$.

\subsection{MDP Reformulation of Problem $\mathbf{P}1$}

We define the problem $\mathrm{P} 1$ as an infinite-horizon Markov decision process (MDP) by defining a tuple with five elements ${\mathcal{S}^p,\mathcal{A}^p,\mathcal{P}^p,\mathcal{R}^p,\gamma^p}$. Here, $\mathcal{S}^p$ is the state set, $\mathcal{A}^p$ is the action set, $\mathcal{P}^p$ is the transition probability, $\mathcal{R}^p$ is the reward function, and $\gamma^p$ is the discount factor $(0 < \gamma ^p< 1)$. The MDP is defined as follows,
\subsubsection{\textbf{State}} In each decision step $i$, we assume that the decision step is in time slot $t$, and $\mathcal{K}_i \subset \mathcal{K}_t$ represents the set of frames waiting for transmission. The state is represented by $\bm{s}_i^p=\{\bm{s}_{n,k,i}^p \ | \ \forall (n,k) \in \mathcal{K}_i\}$, where $\bm{s}_{n,k,i}^p=(w_{n,k,i},\widetilde{r}_{n,k,i},\tau_{n,k,i},{c}_{n,i},N^{rb}_i)$ consists of the data size waiting for transmission $\widetilde{r}_{n,k,i}=\widetilde{r}_{n,k,t} $, the RFDB $\tau_{n,k,i}=\tau_{n,k,t}$, the achievable data rate of each RB ${c}_{n,i}={c}_{n,t}$, the number of remaining RBs $N^{rb}_i$, and the frame importance weight $w_{n,k,i}=w_{n,k}$. Therefore, the state $\bm{s}_i^p$ can be represented as a state matrix of size $(|\mathcal{K}_i|,5)$.

\subsubsection{\textbf{Action}} In each decision step $i$, the agent selects one frame data to transmit. The action is given by $\bm{a}_i^p=(n_i,k_i) \in \mathcal{A}_i^p =\mathcal{K}_i$. Given an action, the number of RBs allocated to the selected frame is $\mathop{\textrm{min}} ( \lceil \frac{ \widetilde{r}_{n,k,i}}{c_{n,i}} \rceil,N_i^{rb})$.

\subsubsection{\textbf{Transition Probability}} Given the current state $\bm{s}_i$ and action $\bm{a}_i$, the transition probability $P(\bm{s}_{i+1}^p|\bm{s}_i^p,\bm{a}_i^p)$ can be defined. As depicted in Fig. \ref{fig-priority}, there are three types of state transitions.
\begin{itemize}
  \item{\textit{Type 1: Successful Transmission.}} With the chosen action, one frame is successfully transmitted regardless of whether RBs are used up.
  \item{\textit{Type 2: Resource Exhaustion}}. With the chosen action, RBs are used up, while the selected frame is not successfully transmitted.
  \item{\textit{Type 3: Slot Transition}}. When RBs are used up or the frame set is empty, the state transitions to a new state of the new slot by updating $w_{n,k,i}$, $\tau_{n,k,i}$, ${c}_{n,t}$, $N^{rb}_i$, and the frame set $K_i$ (discard frames with RFDB $\tau_{n,k,i}=1$ and add arrived frames). 
\end{itemize}
For type 1 and type 2, the transition probability $P(\bm{s}_{i+1}^p|\bm{s}_i^p,\bm{a}_i^p)$ is 1. For type 3, the transition probability $P(\bm{s}_{i+1}^p|\bm{s}_i^p,\bm{a}_i^p)$ depends on the frame arrival distribution, and the action does not affect the state transition.

\subsubsection{\textbf{Reward}} Under the selected action $\bm{a}^p_i=(n_i,k_i)$ at the decision step $i$, the reward function can be obtained based on the types of state transitions. Let $\widetilde{\mathcal{K}}_{n,t,i}$ denote the set of dropped frames
at the decision step $i$. Therefore, we formulate the reward function as,
\begin{eqnarray}\label{equ3-B-2}  
r^p(\bm{s}_i^p,\bm{a}_i^p)=\!  \left\{
{\begin{aligned}
0  \qquad \qquad \qquad \quad  , \  \mathop{\textrm{If type 1 or type 2}}, \\
   -\sum_{n \in \mathcal{N}} \! \sum_{k \in \widetilde{\mathcal{K}}_{n,t,i}} \!\!\!\! w_{n,k,i}, \qquad \qquad  \ \mathop{\textrm{If type 3}}. \\
\end{aligned}}  \right.
\end{eqnarray}

In our problem, the agent selects the highest priority frame for transmission. This policy determines the probability of the agent choosing each action at a given state, and can be expressed as $\pi^p(\bm{a}_i^p,\bm{s}_i^p)$. The criterion for selecting an action based on priority is similar to the policy used in MDP.

Given a policy $\pi$, the on-policy value function is defined as the expected total discounted reward.
\begin{eqnarray}\label{equ3-A-1}  V^p_{\pi}(\bm{s}^p)=\mathbb{E}_{\bm{a}_i^p\sim{\pi(\cdot|\bm{s}_i^p)} } [\sum_{i=0}^{+ \infty} (\gamma^p)^{i}r^p(\bm{s}_i^p,\bm{a}_i^p) |\bm{s}_0^p=\bm{s}^p].
\end{eqnarray}
The corresponding on-policy action-value function can be given by,
\begin{eqnarray}\label{on_policy_q} 
Q_{\pi}^p(\bm{s}^p, \! \bm{a}^p)
\! \! \! \! &=& \! \! \! \! \! \mathbb{E}^{\bm{a}_i^p\sim{\pi(\cdot|\bm{s}_i^p)}}_{\bm{s}_{i+1}^p\sim{P(\cdot|\bm{s}_i^p,\bm{a}_i^p)} } [\sum_{i=0}^{+ \infty} \! \! \gamma^{i}r(\bm{s}_i^p,\bm{a}_i^p) |\! \bm{s}^p_0=\bm{s}^p,\! \bm{a}^p_0=\bm{a}^p] \nonumber \\
\! \! \! \! &=&\! \! \! \! \! \mathbb{E}_{\bm{s}_{i+1}^p\sim{P(\cdot|\bm{s}_i^p,\bm{a}_i^p)} } [r(\bm{s}_i^p,\bm{a}_i^p) \nonumber\\ 
\! \! \! \! & & \! \! \! \! \! +\gamma\mathbb{E}_{{\bm{a}_{i+1}^p\sim{\pi(\cdot|\bm{s}_{i+1}^p)}}} [Q_{\pi}(\bm{s}_{i+1}^p,\bm{a}_{i+1}^p)]] 
\end{eqnarray}
According to \eqref{on_policy_q}, when the policy $\pi$ is different, the action-value function is also different. We define the optimal action-value function as $ Q_{\pi}^{p*}(\bm{s}^p,\bm{a}^p)=\mathop{\textrm{max}} _{\pi}  Q_{\pi}^p(\bm{s}^p,\bm{a}^p)$. Based on $ Q_{\pi}^{p*}(\bm{s}^p,\bm{a}^p)$, the optimal policy can be given by, 
\begin{eqnarray}\label{equ3-B-2}  
\pi^{p*}(\bm{s}_i^p,\bm{a}_i^p)=\!  \left\{
{\begin{aligned}
1  \ , \  \mathop{\textrm{if}} Q_{\pi}^p(\bm{s}_i^p,\bm{a}_i^p)=\mathop{\textrm{max}}_{\bm{a}' \in \mathcal{A}_i^p}  Q_{\pi}^{p*}(\bm{s}^p_i,\bm{a}')  \\
0 \ , \qquad \qquad   \ \mathop{\textrm{otherwise}} \qquad \qquad \qquad ,
\end{aligned}}  \right.
\end{eqnarray}
Therefore, the Bellman equation for the optimal action-value function is, 
\begin{eqnarray}\label{Bellman_optimal_action_value}  Q_{\pi}^{p*}(\bm{s}^p_i,\bm{a}^p_i) = \mathbb{E}_{\bm{s}_{i+1}^p\sim{P(\cdot|\bm{s}_i^p,\bm{a}_i^p)} } [r(\bm{s}_i^p,\bm{a}_i^p) \nonumber \\ +\gamma \! \! \mathop{\textrm{max}}_{\bm{a}' \in \mathcal{A}_{i+1}^p} \! \! Q^{p*}(\bm{s}_{i+1}^p,\bm{a}')]. \nonumber
\end{eqnarray}

However, the optimal action-value function can not be obtained directly, because of the lack of prior information on transition probabilities $P(\bm{s}_{i+1}^p|\bm{s}_i^p,\bm{a}_i^p)$. A Q-learning method based on off-policy temporal difference is proposed to estimate the optimal action-value function $Q^*(s_i,a_i)$ iteratively by, 
\begin{eqnarray}\label{Qlearning}  Q^p(\bm{s}_i^p,\bm{a}_i^p) \! \! \! \! &=& \! \! \! \! Q^p(\bm{s}_i^p,\bm{a}_i^p)+\alpha[ r^p(\bm{s}_i^p,\bm{a}_i^p) \nonumber\\ 
\! \! \! \! & & \! \! \! \! + \gamma^p  \! \! \mathop{\textrm{max}}_{\bm{a}'\in \mathcal{A}_{i+1}^p}  \! \! Q(\bm{s}_{i+1}^p,\bm{a}')-Q^p(\bm{s}_i^p,\bm{a}_i^p)],
\end{eqnarray}

In order to deal with the challenge of the large state space, DQN is proposed. The neural network of DQN is parameterized by weights and biases denoted as $\theta^p$. To estimate the neural network, we can optimize the following loss functions at iteration $j$,
\begin{eqnarray}\label{Bellman_error}
L(\theta_j^p)=\mathbb{E}\bigg[\bigg(y(\hat{\theta}^p_j)-Q^p(\bm{s}_i^p,\bm{a}_i^p;\theta_j^p)\bigg)^2\bigg],
\end{eqnarray}
where
\begin{eqnarray}\label{equ:target}
y(\hat{\theta}^p)=r^p(\bm{s}_{i}^p,\bm{a}_{i}^p)+ \! \! \gamma^p \mathop{\textrm{max}}_{\bm{a}' \in \mathcal{A}_{i+1}^p}  \! \! Q^p(\bm{s}_{i+1}^p,\bm{a}';\hat{\theta}^p_j),
\end{eqnarray}
and $\hat{\theta}^p_j$ is the parameter of a fixed target network at iteration $j$ \cite{cite:Dueling}. Therefore, the specific gradient update can be given by,
\begin{eqnarray}\label{Bellman_error}
\nabla_{\theta_j^p}L(\theta_j^p)=\mathbb{E}\bigg[\bigg(y(\hat{\theta}^p_j)-Q(\bm{s}_i^p,\bm{a}_i^p;\theta_j^p)\bigg)\nabla _{\theta_j^p}Q(\bm{s}_i^p,\bm{a}_i^p;\theta_j^p)\bigg], \nonumber 
\end{eqnarray}

\subsection{DQN Model Design and Training}

\begin{figure}[tb]
\centering 
\includegraphics[height=1.6in,width=3.5in]{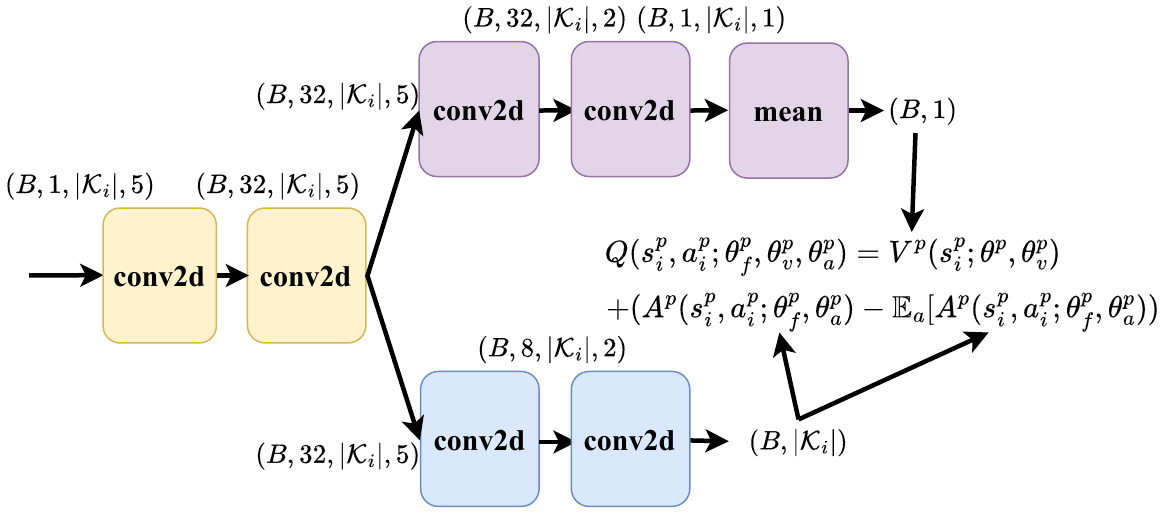}
\caption{CNN-based Neural network design for MS-DQN. This model can be applied to state and policy spaces of different sizes.}
\label{fig:NN} 
\vspace{-3mm}
\end{figure}

Two improvements were made to the original algorithm to develop an algorithm that is tailored to this task. Firstly, as previously mentioned,  there are three types of state transitions $\pi^p(\bm{a}_i^p,\bm{s}_i^p)$. In types 1 and 2, rewards depend on both states and actions, while in type 3, they depend only on states.  In order to better estimate the rewards brought by states and actions, we decouple the action-value function into a value function and an advantage function that quantifies the advantage of each action. More specifically, we have,
\begin{eqnarray}\label{Bellman_error}
Q^p(\bm{s}_{i}^p,\bm{a}_{i}^p)=V^p(\bm{s}_{i}^p)+A^p(\bm{s}_{i}^p,\bm{a}_{i}^p).
\end{eqnarray}
where $A^p(\bm{s}_{i}^p,\bm{a}_{i}^p)$ denotes the advantage function. To implement this decoupling, we use the Dueling DQN method \cite{cite:Dueling}, and design the last module of the network as follows,
\begin{eqnarray}\label{Bellman_error}
Q^p(\bm{s}_i^p,\bm{a}_i^p;\theta_f^p,\theta_v^p,\theta_a^p)  = V^p(\bm{s}_i^p;\theta_f^p,\theta_v^p)+ \qquad  \qquad \quad  \nonumber \\ 
\qquad \qquad \qquad \quad (A^p(\bm{s}_i^p,\bm{a}_i^p;\theta_f^p,\theta_a^p) 
-\mathbb{E}_a[A(\bm{s}_i^p,\bm{a}_i^p;\theta_f^p,\theta_a^p)
]), \nonumber
\end{eqnarray}
where $\theta_f^p$, $\theta_v^p$ and $\theta_a^p$ are parameters of feature extraction network, value network, and action network, respectively. 

Another challenge is that the input and output sizes of the network are not fixed. As mentioned earlier, the input size of the network is $(B,1,|\mathcal{K}_i|,5)$, and the output size is $(B,|\mathcal{K}_i|)$, where $B$ is the batch size. However, $|\mathcal{K}_i|$ can have different values in different states. The size of the action space and state space is dynamically changing. To overcome this, one solution is to use an upper bound $K$ for $|\mathcal{K}_i|$ (where $K \ge \mathcal{K}_i$) and pad the remaining input parameters with zeros \cite{cite:DDPG-padding}, or train different models for different input sizes directly \cite{cite:DDPGDNN}. These model design methods are not always general enough to be deployed in real systems.

Combining these two aspects, we designed our model based on 2D-CNN layers, as shown in Fig. \ref{fig:NN}. There are two advantages of using CNN. Firstly, CNN can adapt to parameter inputs of different dimensions, making it suitable for variable dimensional learning tasks. Secondly, CNN has low computational complexity (can perform parallel computations), and the computational complexity increases linearly with the number of users, ensuring real-time operation of the algorithm.

In each decision step, we store the agent's experience $(\bm{s}_i,\bm{a}_i,\bm{s}_{i+1},\bm{r}(a_i,s_i))$ in the replay buffer. When training the neural network, we sample data in batches from the replay buffer. We train the network using the $\epsilon$-greedy policy \cite{cite:Dueling}. Specifically, when the state transitions to type 3 and $\mathcal{K}_i=\emptyset$, we set the action to a default action $\bm{a}_i=(0,0)$, and let $Q(\bm{s}_i,\bm{a}_i)=0$.

\section{EXPERIMENTAL RESULTS } \label{Sec5}

In this section, we evaluate the performance of the proposed algorithm. For all results, unless specified otherwise, the bandwidth is 50MHz with sub-carrier spacing (SCS) 30KHz. We consider the Urban Microcell deployment scenario. Devices are distributed in the area within [500m, 500m] randomly. In our system, the bitrate of the XR video is assumed to be 20Mbps, and the frame rate is 60FPS. Therefore, the mean size of each frame is 250Kbits. The jitter of frame size and arrival time is generated according to \cite{cite:TS38838}. I-frames are transmitted periodically every $K=4$ frames, while the other frames are P-frames. The average size ratio between one I-frame and one P-frame is $\alpha=1.5$. The FDB is set to be 10ms. Results are obtained through a 2000-slot simulation. The distribution of the first frame's arrival time can be represented as $t_{n,0} \in [0,1000/f_n]$ ms. The importance weights of I-frames and P-frames are set to 1 and 0.1, respectively. It should be noted that the importance of I-frames and P-frames can be set according to actual needs. The algorithm we proposed is effective under different importance settings.

To compare BS scheduling schemes, we fix the bitrate of XR video and compare the proposed MS-DQN algorithm with baseline scheduling algorithms. We consider the following baseline algorithms.

\subsubsection{\textbf{DDPG\cite{cite:priority-DDPG}}} The DDPG algorithm scheduler uses a neural network to adjust the off-policy Deterministic Policy Gradient algorithm based on an Actor-Critic framework. 
\subsubsection{\textbf{PF\cite{PF}}} The PF scheduler considers the device's current rate and historical average rate to calculate the priority.
\subsubsection{\textbf{PF-I\cite{cite:frameintegrated}}} The Frame-level integrated based PF scheduler is designed to take into account both the integrity of the frame and the network channel conditions for devices.

\subsection{Comparison of Convergence Performance} 

\begin{figure}[tb]
\centering 
\includegraphics[height=1.9in,width=2.5in]{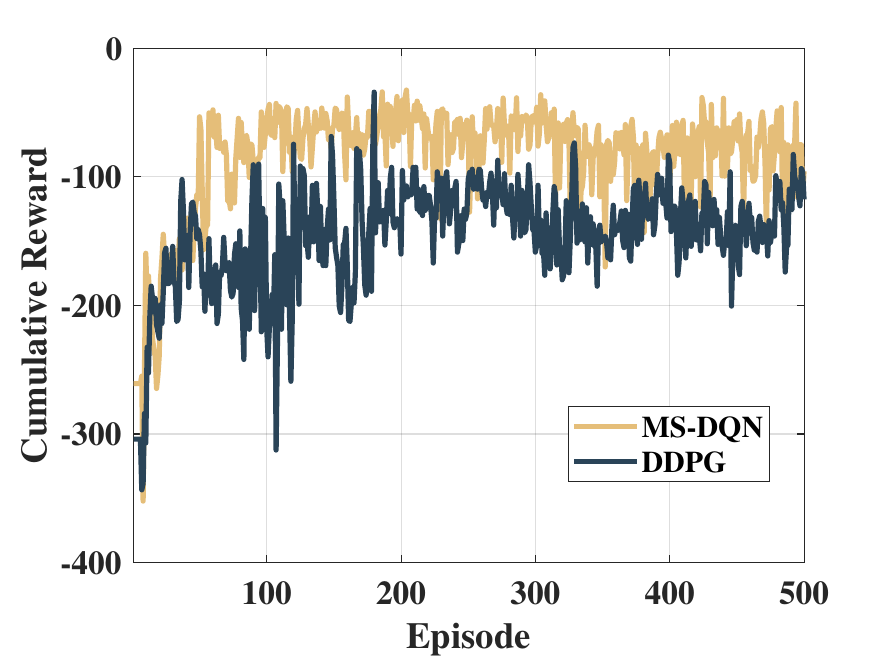}
\caption{Convergence performance of MS-DQN and DDPG algorithms.}
\label{fig:TransCov} 
\vspace{-5mm}
\end{figure}

We first compare the convergence performance between our proposed MS-DQN algorithm and the DDPG algorithm However, when the number of devices and the initial time of the first frame are different, the convergence curve will also be different. For the fairness of the comparison, in each episode, we train the neural network using scenarios with varying numbers of devices and initial times and test with the same scenario to observe changes of cumulative reward. Each episode of training and testing is 2000 slots. Fig. \ref{fig:TransCov} depicts the cumulative reward of different algorithms. The number of devices is 8. Both the proposed MS-DQN-based algorithm and the DDPG algorithm can converge after about 50 episodes. However, compared with the DDPG algorithm, the MS-DQN algorithm's cumulative reward is higher, which shows the proposed MS-DQN algorithm performs better than DDPG. 

\subsection{Performance Comparison}

\begin{figure}[tb]
\centering 
\includegraphics[height=2.0in,width=2.5in]{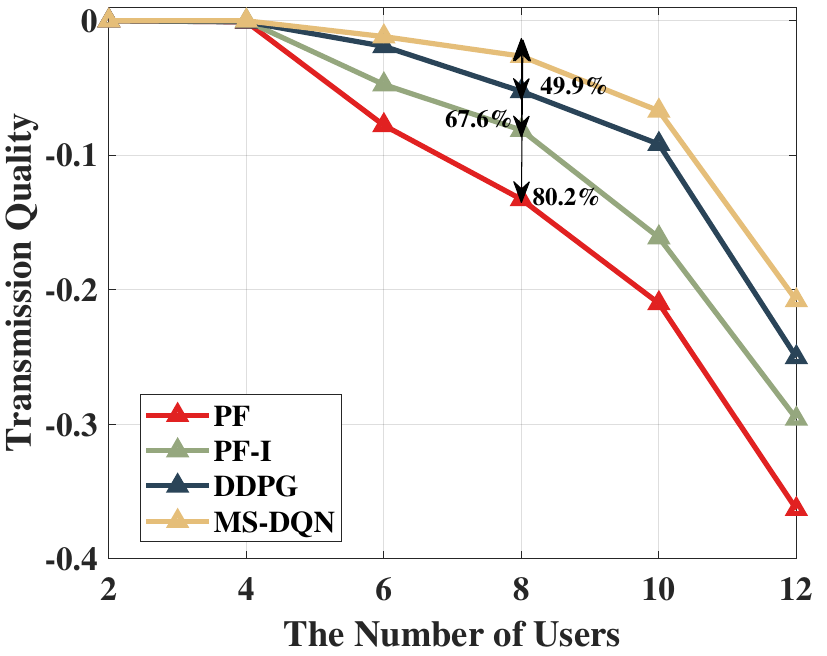}
\caption{Performance comparison of transmission quality between the proposed MS-DQN algorithm and baseline algorithms.}
\label{fig:TransQualityResult} 
\vspace{-3mm}
\end{figure}

Fig. \ref{fig:TransQualityResult} shows the comparison of transmission quality among different algorithms under different numbers of devices. We repeated the experiment 50 times, and each experiment ran a simulation of 2000 slots. The proposed MS-DQN algorithm outperforms the baseline algorithm. For example, when the number of devices is 8, compared with PF, PF-I, and DDPG algorithms, MS-DQN has an improvement of 80.2\%, 67.6\%, and 49.9\% in transmission quality, respectively.

\begin{figure}[tb]
\centering 
\includegraphics[height=2.0in,width=2.8in]{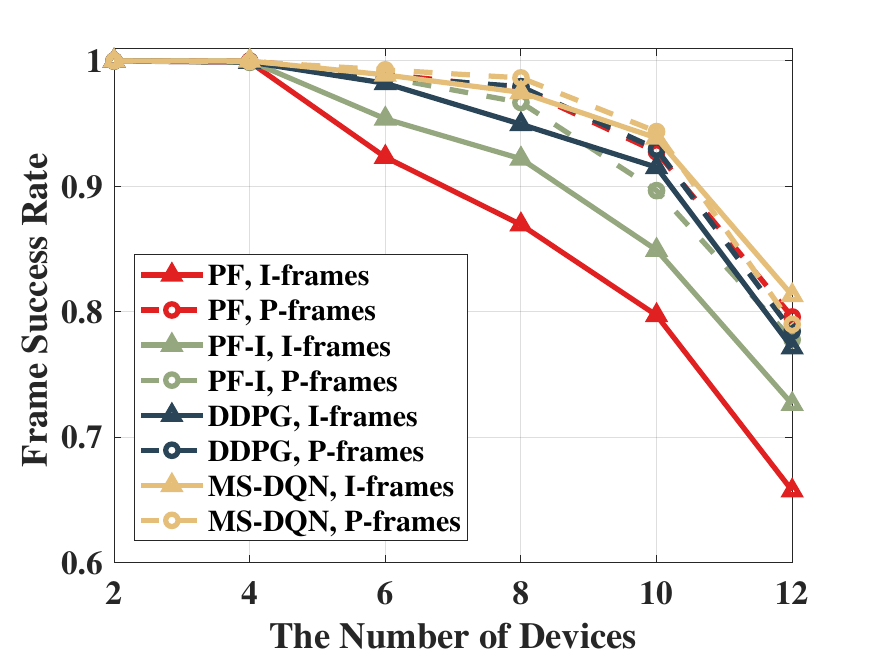}
\caption{I-frames and P-frames transmission performance comparison between the proposed MS-DQN algorithm and baseline algorithms.}
\label{fig:TransQualityResultIPFrame} 
\vspace{-3mm}
\end{figure}

 Fig. \ref{fig:TransQualityResultIPFrame} illustrates the impact of different schemes on the frame success rates of I-frames and P-frames. Specifically, for I-frame transmission, compared to the PF, PF-I, and DDPG schemes, the proposed MS-DQN scheme can improve the frame success rate by 10.6\%, 5.3\%, and 2.5\% respectively when the number of devices is 8. For P-frame transmission, MS-DQN achieves a frame success rate improvement of 0.7-2.0\% when the number of users is 8.

\begin{figure}[tb]
\centering 
\includegraphics[height=2.0in,width=2.5in]{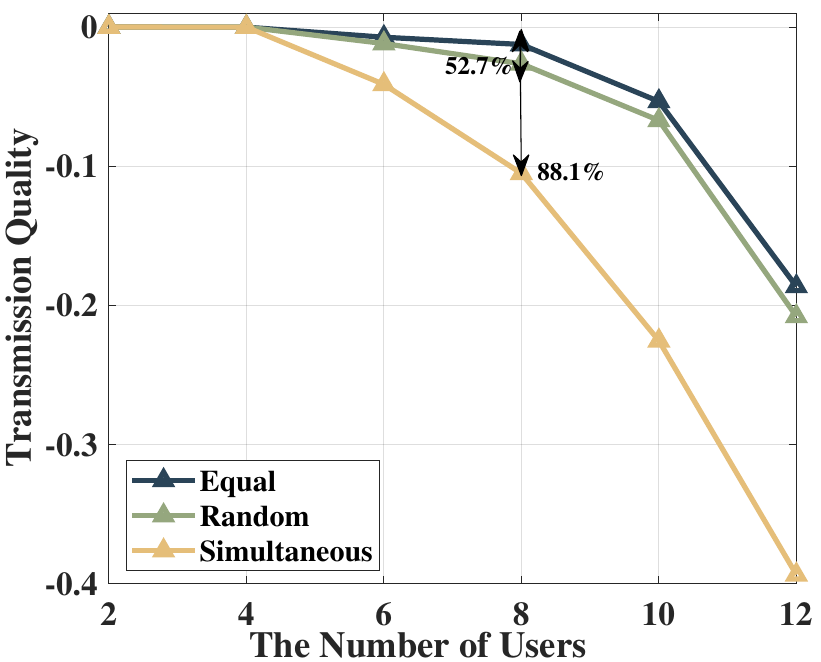}
\caption{Performance comparison of transmission quality with different initial arrival times.}
\label{fig:TransQualityResultSYN} 
\end{figure}

Considering the impact of the initial arrival time, we tested three different scenarios. Scenario 1: the initial frame arrival follows a uniform distribution. (Random); Scenario 2: the initial frames of all devices arrive simultaneously (Simultaneous); Scenario 3: the initial frames of all devices arrive with equal intervals (Equal). As shown in Fig. \ref{fig:TransQualityResultSYN}, when the number of devices is 8, compared with the Random and Simultaneous settings, the Equal setting can achieve an 88.1\% and 52.7\% increase in transmission quality, respectively. This indicates that the frame arrival interval of different devices can affect the frame success rate of BS transmission. XR video servers can use this feature to arrange and adjust encoders to optimize real-time XR video transmission.

\section{Conclusion} \label{Sec6}

In this paper, we consider optimizing XR transmission in the BS. We propose a scheduling method based on frame priority and design a scheduling framework based on MS-DQN. Based on system modeling and problem requirements, we designed a multi-step scheduling strategy and a CNN-based neural network model to address the issues of frame-by-frame scheduling and dynamic changes in the size of state space and action space. Experiments show that our proposed MS-DQN framework and frame-priority-based scheduling method can effectively improve transmission quality. Compared with the baseline algorithms, our proposed algorithm can increase the transmission quality by 49.9\%-80.2\%.

\section*{ACKNOWLEDGEMENT}
This work was supported in part by the National High Quality Program grant TC220H07D, the National Natural Science Foundation of China (NSFC) under Grant 61871262, 62071284, and 61901251, the National Key R\&D Program of China under Grant 2022YFB2902005 and 2022YFB2902000, Key-Area Research and Development Program of Guangdong Province grant 2020B0101130012, Foshan Science and Technology Innovation Team Project grant FS0AAKJ919-4402-0060.

\bibliographystyle{IEEEtran}

\bibliography{reference}

\end{document}